\documentstyle[12pt]{article}

\begin{document}
\input{amssym.def}

\newcommand{\vp}{\varphi}
\newcommand{\br}{{\Bbb R}}
\newcommand{\bt}{{\Bbb T}}
\newcommand{\bz}{{\Bbb Z}}
\newcommand{\bn}{{\Bbb N}}
\newcommand{\bc}{{\Bbb C}}
\newcommand{\ct}{{\cal T}}
\newcommand{\lr}{\longrightarrow}
\newcommand{\ra}{\rightarrow}
\newcommand{\cg}{{\cal G}}
\newcommand{\cc}{{\cal C}}
\newcommand{\ol}{\overline}
\newcommand{\bfm}{{\bf m}}
\newcommand{\bfn}{{\bf n}}
\newcommand{\ds}{\displaystyle}
\newcommand{\cu}{{\cal U}}
\newcommand{\ch}{{\cal H}}
\newcommand{\ca}{{\cal A}}
\newcommand{\cb}{{\cal B}}
\newcommand{\cs}{{\cal S}}
\newcommand{\wt}{\widetilde}
\newcommand{\wh}{\widehat}

\title{\Large \bf Heisenberg Groups in the Theory of the Lattice
Peierls Electron: the Irrational Flux Case}
\author{~~~~ \\ [.36in]
P.P. Divakaran \\
Chennai Mathematical Institute\thanks{Work supported by the
Department of Science Technology, Government of India.} \\
92 G.N. Chetty Road, T. Nagar \\
Chennai-600 017, India.\\
E-mail: ppd@smi.ernet.in
}
\date{}
\maketitle

\newpage 

\begin{center}
{\sf Abstract}
\end{center}

\vspace{2mm}
This paper establishes that the quantum mechanics of a
charged particle moving in $\cc = \br^2$ (Landau) or $\bz^2
\subset \br^2$ (Peierls) in a uniform normal magnetic field
$B$ is described in every detail by the (projective)
representation theory of the appropriate euclidean group
$E(2) = J \vec{\times} A^2$ ($\vec{\times}$ is the semidirect product;
$J = SO(2, \br )$, $A = \br$ and $J = \bz/4$, $A = \bz$ for
$\cc = \br^2$ and $\bz^2$ respectively).  The central
extensions of $E(2)$ by the circle group $\bt$ are of the form
$\wt{E}(2) = J \vec{\times} \wt{A}^2$ and hence the main object of
study is the nilpotent group $\wt{A}^2$.  The unique
representation property of the Heisenberg group
$\wt{\br}^2_r$, $r \in H^2 (\br^2, \bt ) = \br$
leads to a detailed
description of the structure of the state space $\ch_r \cong
L^2 (\br^2)$ as a representation of $\wt{\br}_r^2$ and of
$\wt{E}_r(2,\br )$.  An essentially unique Hamiltonian $H_r$
for each $\wt{E}_r (2,\br )$ is also determined.  The
quantum theory that results is that of the Landau electron
when $\{ B\}$ is identified with $H^2 (\br^2,\bt )$.  For
the Peierls case, the central extensions $\wt{E}_{\theta}
(2,\bz ) = \bz /4 \vec{\times} \wt{\bz}^2_{\theta}$ are
parametrised by $\theta \in [0,2\pi ) \cong \bt$.  When
$\theta /2\pi$ is irrational, $\wt{\bz}_{\theta}^2$ 
is an ``almost Heisenberg'' group in the sense that it has a
distinguished irreducible representation on $L^2(\bz )$.  This is
sufficient for a complete description of $\ch_{\theta} \cong
L^2 (\bz^2)$ as a representation of $\wt{\bz}_{\theta}^2$
and of $\wt{E}_{\theta} (2,\bz )$.  When $\theta$ is
identified with $\Phi$, the flux per plaquette modulo the
flux quantum, the 
physics of the Peierls electron is fully determined by
$\Phi$ and is periodic in $\Phi$ with one flux quantum as
the period.  The Hamiltonian $H_{\theta}$ in $\ch_{\theta}$
is also determined by $\wt{E}_{\theta} (2,\bz )$
invariance; for only nearest neighbour hopping, $H_{\theta}$
is essentially the Harper Hamiltonian.  Introduction of
vector potentials and gauges is nowhere necessary.

\newpage
\noindent
{\sf 1. General Introduction to the Problem and
the Method}

\vspace{2mm}
By the term Peierls electron [1] we mean a quantum system
consisting of a particle of electric charge $e$ and mass $m$ in a
constant uniform magnetic field $B$ and a periodic potential
$V$.  Since motion in the direction of $B$ is independent of
$B$, the configuration space is taken to be the plane
perpendicular to $B$ or any subset of it compatible with
$V$, i.e. if $\Lambda \subset \br^2$ is the lattice defined
by $V , \cc$ should be invariant under automorphisms of
$\Lambda$; in particular if $\cc = \Lambda$, $V$ is constant
on $\cc$ and so can be dropped.  The reason for this
restriction (general discrete subsets of $\br^2$ have been
considered in the literature [2]) is that our method of
treating the problem is the representation theory of the group
of automorphisms of $\cc$.  The specific case which is the
main concern of this paper is the $\bz^2$ lattice Peierls
electron, $\cc = \Lambda = \bz^2$.

In his original study of the continuum problem, $\cc = \br^2,
\Lambda = \bz^2$, Peierls [1] made several perceptive
simplifications of which the one most pertinent to us is the
weak field or tight binding approximation.  Later, Harper
[3] in a careful study found a simple expression for the
Hamiltonian of the electron within the Peierls scheme of
approximations and in the extreme tight binding limit (in
which wave functions are presumably supported in arbitrarily
small neighbourhoods in $\br^2$ of the points of $\Lambda$):
$$H_{\Phi} = p_1 +p_2 +p_1^{-1} + p_2^{-1} \eqno (1)$$
where $p_1$ and $p_2$ are unitary operators on the state
space of the electron satisfying the commutator condition
$$p_1 p_2 p_1^{-1} p_2^{-1} = e^{i\Phi} \eqno (2)$$
and $\Phi$, the only physical parameter left in the model,
is the flux of $B$, in suitable units, through a unit cell
or plaquette.  

In the vast literature on the subject, it is sometimes
claimed that eqn. (1) defines the exact Hamiltonian for the
$\bz^2$ Peierls electron, presumably because of the tight
binding invoked in obtaining it.  To establish rigorously
the validity of such a claim, especially in view of the
other poorly understood approximations made, one would first
need a clear-cut prescription for the quantum mechanics of a
charged particle living on a lattice and subject to a
magnetic field.  A magnetic field on a lattice is a
physically ill-defined concept, a vector potential even more
so.  In practice, what is done is to associate certain
unitary operators constructed from a continuum vector potential to
links between the points of $\Lambda$ in such a way that in
some ``local'' limit, the kinematics and dynamics of the
continuum system are recovered.  This procedure entails
several arbitrary choices, e.g., the choice of a gauge,
which obscure some central issues.  One would like to know
answers to questions such as: What is the exact
characterisation of the state space $\ch$?  How do $p_1$ and
$p_2$ operate on $\ch$ and what is their physical
significance?  To what extent do eqns. (1) and (2) determine
$H_{\Phi}$ as an operator on $\ch$?  How can we tell that
$H_{\Phi}$ is the correct Hamiltonian for the problem and is
it unique? etc.

It is the purpose of this paper and a sequel [4], working
directly with the lattice problem, to show that all such
questions can be 
posed precisely and answered in terms of the (projective)
representations of the group $E$ of automorphisms of the
configuration space ($\bz^2$ in the present case).  The
reason why projective 
representations of Aut$\,\cc$ are so effective in the quantum
theory of magnetic fields is best understood in the example
of the familiar Landau electron, $\cc = \br^2$, $E =
E(2,\br ) = SO(2,\br ) \vec{\times} \br^2$ ($\vec{\times}$ denotes the
semidirect product).  A uniform magnetic field normal to
$\br^2$ is certainly invariant under $E(2,\br )$ and the
quantum theory should be capable of being described by the
representations of $E(2,\br )$.  The linear (unitary)
representations correspond of course to free particle
motion.  However, $E (2,\br )$ has nontrivial projective
representations and since, by Wigner's theorem [5], they are
as legitimate as linear representations, such
representations are the only means of accommodating
$E(2,\br )$  invariant but non-free quantum mechanics.
(Statements to the effect that a uniform
magnetic field violates translation invariance are not rare
in the literature).
Because of the key role of this elementary fact in our
considerations, section 3 is devoted to demonstrating that
the quantum theory of the Landau electron is no more and no
less than the theory of projective representations of
$E(2,\br )$.

The general setting for our approach is provided by the
standard correspondence of the set of equivalence classes of
projective representations (the qualifier ``unitary'' is 
implicit and will be dropped from now on) of a group $E$ with
its 2nd cohomology group $H^2 (E,\bt )$ ($\bt$ is the circle
group) and, equivalently, with the set of isomorphism
classes of central extensions $\{\wt{E}\}$ of $E$ by $\bt$; in
particular, every projective representation of $E$ in the
class of $\alpha \in H^2 (E,\bt )$ lifts to a linear
representation of a central extension $\wt{E}_{\alpha}$
having the property that its restriction to $\bt \subset $
centre $\wt{E}_{\alpha}$ is the natural character $t
\longmapsto t$ (see, for example, [6,7] for these well-known
facts).  The central extensions of $E (2,\br )$ by $\bt$ are
all of the form $\wt{E}_r (2,\br ) \cong SO(2,\br ) \vec{\times}
\wt{\br}_r^2$, $r \in \br$.  The real Heisenberg group $\wt{\br}_r^2$
thus assumes significance, especially its property of
having, upto equivalence, just one irreducible
representation restricting to a fixed character on $\bt$
(the Stone -von Neumann theorem), conveniently realised on
$L^2(\br)$.  The state space of a particle in $\cc = \br^2$
is however isomorphic to $L^2 (\br^2)$ (for the general
theory of the state space of a system defined by $\cc$ and
$E$, see [8,9]).  We 
will study in section 3 the structure of $L^2 (\br^2)$ as a
$\wt{E}_r (2,\br )$ module and see that $L^2 (\br^2) \cong
V_r \otimes V_{-r} = \ch_r$ with $\wt{\br}_r^2$ acting
irreducibly on $V_r$ and trivially on $V_{-r}$ ($V_{-r}$
will turn out to be an irreducible $\wt{\br}_{-r}^2$
module).  We shall then use the above factorisation and a
general characterisation of the Hamiltonian of a quantum
system $(\cc, E)$ [7,8] to determine the Hamiltonian $H_r$
in $\ch_r$.  It will be seen that $H_r$ is precisely the
Landau Hamiltonian (when $r$ is put equal to $eB$)
operating on $V_{-r}$ and having $V_r$ as degeneracy
subspace.  A bonus is that vector potentials and hence
gauges are nowhere required to be invoked.

With the confidence thus acquired (a semi-heuristic account
of the Landau electron from the symmetry point of view can
be found in [10]), we turn to the $\bz^2$ Peirels electron in
section 4.  The relevant symmetry group is the integral
euclidean group $E(2,\bz ) = SO(2,\bz )\vec{\times} \bz^2 = \bz
/4 \vec{\times} \bz^2$.  All central extensions of $E(2,\bz )$ by
$\bt$ are of the form $\wt{E}_{\theta} (2,\bz ) = \bz /4
\vec{\times} \wt{\bz}_{\theta}^2$, parametrised by an angle
$\theta \in \bt = \{ 0 \leq \theta < 2\pi \}$.  However,
$\wt{\bz}_{\theta}^2$ does not have the unique
representation property for any value of $\theta$.  The
modifications in the theory of Heisenberg groups necessary to
deal with the situation are described in section 2.  The
distinctive features are as follows.

A central extension $\wt{G}$ of an abelian group $G$ by
$\bt$ is uniquely characterised by the function $c: G \times
G \lr \bt$ induced by the commutator map $(\wt{g}, \wt{h})
\longmapsto \wt{g} \wt{h} \wt{g}^{-1} \wt{h}^{-1}$ in
$\wt{G}$ [11].  Denote the Pontryagin dual of $G$ by
$\wh{G}$ and associate to every $\wt{G}$ a homomorphism $\mu
: G \lr \wh{G}$ by $(\mu (g))(h) = c(g,h)$.  Following
Mumford [12], we call a $\wt{G}$ for which $\mu$ is an
isomorphism a Heisenberg group as every such $\wt{G}$ has
only one irreducible representation (upto equivalence) such
that it restricts to a fixed character on $\bt$.  

Since $\bz^2$ is not self-dual, $\wt{\bz}_{\theta}^2$ cannot
be Heisenberg for any $\theta$; depending on the value of
$\theta$, it belongs to one of two types of generalisations
of Heisenberg groups:
\begin{itemize}
\item If $\theta$ is an irrational multiple of $2\pi$,
$\wt{\bz}^2_{\theta}$ is a dense subgroup of a Heisenberg
group; it has then a distinguished irreducible
representation obtained by restriction.
\item If $\theta = 2\pi \nu /N$, $\nu$ and $N$ coprime,
$\wt{\bz}_{\theta}^2$ is a central extension by $(N\bz
)^2$ of the finite Heisenberg group $(\bz
/N)^{2\sim}_{\theta}$.  Inequivalent irreducible
representations of $\wt{Z}_{\theta}^2$ are classified by the
characters of $(N\bz )^2$.
\end{itemize}

In both cases, the relevant representations of $\wt{E}_{\theta}
(2,\bz )$ provide a full description of the $\bz^2$ Peierls
electron in flux per plaquette $\Phi$ when $\theta$ is
identified with $\Phi$ (mod $\bz$) in units of the flux
quantum $2 \pi /e$.  The present paper is, however, confined
to the irrational flux case (the rational case will be
covered in a sequel [4]).  In many ways, the theory has parallels
with the Landau electron; in particular, the state space has the
decomposition $\ch_{\theta} \cong L^2 (\bz^2) \cong
V_{\theta} \otimes V_{-\theta}$ with $\wt{\bz}_{\theta}^2$
acting irreducibly on $V_{\theta}$ and trivially on
$V_{-\theta}$.  The physics is controlled by the flux $\Phi$
and is periodic in $\Phi$ with period 1.  We also determine
all possible Hamiltonians as self-adjoint operators on
$\ch_{\theta}$ invariant under $\wt{E}_{\theta}$ $(2,\bz )$
and find that the simplest of them has the Harper form,
acting on $V_{-\theta}$ and having $V_{\theta}$ as
degeneracy subspace.

Some of these results may come as a surprise to those who
are familiar with the extensive literature on what are
called magnetic translation groups (for irrational fluxes);
see, for instance, [13] and the references cited there.
While it is true that $\wt{\bz}_{\theta}^2$ for $\theta$
irrational has a rich collection of inequivalent
representations, known also from the theory of irrational
rotation $C^*$ algebras [14,15], the one relevant for
quantum mechanics is uniquely given as inherited from an
embedding Heisenberg group.

The mathematical material, on Heisenberg groups and their
appropriate generalisation, is gathered together in section
2.  It is conceptually self-sufficient, though proofs are
not always given in full detail.  The omitted
measure-theoretic and analytic elaborations are standard and
can be supplied without difficulty by the reader.

\vspace{2mm}
\noindent
{\sf 2. Heisenberg Groups and Almost Heisenberg Groups}

\vspace{2mm}
Let $G$ be an abelian group and $\bt$ the circle group, both
written multiplicatively ($\bt \cong U(1)$) and $\wt{G}$ a
central extension of $G$ by $\bt$.  The commutator map in
$\wt{G}, (\wt{g},\wt{h}) \longmapsto
\wt{g}\wt{h}\wt{g}^{-1}\wt{h}^{-1}$, associates to $\wt{G}$
a function $c: G \times G \lr \bt$ which is homomorphic in
each argument and is alternating: $c(g,g) =1$ for all $g \in
G$.  If $\gamma$ is any 2-cocycle on $G$ associated to
$\wt{G}$, then $\gamma (g,h) \gamma (h,g)^{-1} = c(g,h)$ for
all $g,h \in G$.  So changing $\gamma$ by a coboundary does not
affect $c$.  Thus to every central extension of $G$ by $\bt$
corresponds a unique element of the abelian group $\ca^2
(G)$ of alternating bihomomorphic maps (or bicharacters) $G
\times G \lr \bt$.  These basic facts are easy to verify.  

It is less trivial to establish that this correspondence is
in fact bijective [11] (see also [16]):

\noindent
{\bf 2.1.}
{\it $\ca^2(G)$ is isomorphic to the 2nd cohomology
group $H^2 (G,\bt )$}.

Let $\cb^2 (G)$ be the group of all bicharacters of $G$ 
and $\cs^2 (G)$ its subgroup of symmetric bicharacters.
Then $\ca^2 (G) = \cb^2 (G)/\cs^2 (G)$.  In general,
however, $\cb^2 (G) \neq \ca^2 (G) \times \cs^2 (G)$;
$c \in \ca^2 (G)$ is not necessarily skew symmetric, i.e., does
not satisfy $c(g,h)c(h,g)=1$.  A sufficient condition for
alternating to imply skewsymmetric is for $g \longmapsto
g^2$ to be an automorphism of $G$ so that we can define the
square root $g^{1/2}$ of every $g \in G$ as the inverse
of $g \longmapsto g^2$.  If this condition is met,
given any $c \in \ca^2 (G)$, define $\gamma \in \ca^2(G)$ by
$\gamma (g,h) = c(g^{1/2},h) = c(g,h)^{1/2}$ so that $c(g,h)
= \gamma (g,h)^2 = \gamma (g,h) \gamma (h,g)^{-1}$.  This
gives us a canonical 2-cocycle which is itself skewsymmetric
(and hence alternating) for every $\wt{G}$.  These points
are explained in [11].  But, having raised them, for the reason
that $\bz^2$ does not meet the sufficient condition of being
divisible by 2 (in the usual terminology appropriate for
additively written groups) we shall henceforth ignore them;
for irrational central extensions $\wt{\bz}^2$ which concern
us here, it is always possible to choose a skewsymmetric
2-cocycle as will be seen below.

Given $\wt{G}$ and the associated bicharacter $c \in \ca^2
(G)$, denote the Pontryagin dual of $G$ by $\wh{G}$ and
define, following [12], a homomorphism $\mu: G \lr \wh{G}$
by $(\mu (g))(h) = c(g,h)$.  The map $\mu$ decides when
$\wt{G}$ has the unique representation property [12]:

\noindent
{\bf 2.2.}
{\it If $\wt{G}$ is such that $\mu$ is an isomorphism
of $G$ and $\wh{G}$, then all irreducible representations of
$\wt{G}$ which restrict to $\bt \subset $ centre $\wt{G}$ as
the natural character $t \longmapsto t$ are equivalent
representations.}

A central extension by $\bt$ of an abelian group $G$ for
which $\mu$ is an isomorphism of $G$ and $\wh{G}$ is a {\it
Heisenberg extension of $G$} or, simply, a {\it Heisenberg
group}. 

It is known [12] that equivalent realisations of the unique
representation of a Heisenberg group are classified by the
maximal isotropic subgroups, i.e., maximal subgroups $H$ of
$G$, over which $\wt{G}$ splits (so $H$ is a subgroup
of $\wt{G}$): for any such $H \subset G$, there is an action
$U$ of $\wt{G}$ on $L^2 (G/H)$ which is linear (unitary) and
irreducible.  If $G$ is of the form $G = A \times A$ with
$A$ self-dual, we may make the choice $H= A \times
1$.  Writing $\wt{g} \in \wt{G}$ as $(a_1,a_2,t)$ with 
$(a_1,a_2) \in A \times A$, $t \in \bt$, the corresponding
representation on $L^2 (1 \times A)$ can be given as
$$(U(1,1,t)f)(x) = tf(x), \eqno (3)$$
$$(U(a_1,1,1)f)(x) = c((a_1,1),(1,x))f(x), \eqno (4)$$
$$(U(1,a_2,1)f)(x) = f(a_2x). \eqno (5)$$
We remark that the essential reason why this representation
is irreducible is that $c$ is nondegenerate -- i.e., there
exists no $g \in G$, $g \neq Id$, such that $c(g,h) = 1$ for
all $h \in G$ -- which follows from $\mu$ being an
isomorphism. 

Suppose now that $G$ is a non-self-dual group of the form $G
= A^2$ (so $A$ is also not self-dual) and $\wt{G}$ a central
extension of $G$ by $\bt$.  $\wt{G}$ still defines a unique
bicharacter of $G$, $c \in \ca^2(G)$, and a homomorphism
$\mu : G \lr \wh{G}$ as earlier, but $\mu$ cannot be an
isomorphism.  What we demand of $\mu$ now 
is that it should be injective and that its image should be
dense in $\wh{G}$.  We shall call a $\wt{G}$ for which $\mu$
has the above property an {\it almost Heisenberg} group.
Again as before, $A \times 1$ is an isotropic subgroup of
$\wt{G} : c(a,b) = (\mu (a))(b)=1$ for all $a, b \in A
\times 1$. This means that, as maps from $G$ into $\bt$, $\{
\mu (a) \mid a \in A \times 1\}$ have $A \times 1$ as kernel
and hence define maps from $G/(A\times 1)=1\times A$ into
$\bt$.  In other words, the restriction of $\mu$ to $A
\times 1$ maps it into $1 \times \wh{A}$ and has, by
hypothesis, a dense image in $1 \times \wh{A}$.  We have
thus a dense inclusion of $G = A \times A$ in the self-dual
group $G^* = A \times \wh{A}$ by $(a_1,a_2) \longmapsto
(a_2,\mu (a_1))$.

For any almost Heisenberg extension $\wt{G}$ of $G = A^2$,
define a map $c^*: G^* \times G^* \lr \bt$ by $c^* ((a_2,\mu
(a_1)),(b_2,\,\mu (b_1))= c((a_1,a_2),(b_1,b_2))$.  $c^*$ is
an alternating bicharacter defined, to begin with, on a
dense subgroup of $G^* \times G^*$ and, by continuity, on
all of $G^* \times G^*$.  Correspondingly, we have a central
extension $\wt{G}^*$ of $G^*$ by $\bt$.  Thus 

\noindent
{\bf 2.4.}
{\it Suppose $G = A\times A$ is not self-dual and let
$\wt{G}$ be an almost Heisenberg extension of $G$.  Then
there exists a Heisenberg extension $\wt{G}^*$ of $G^* = A
\times \wh{A}$ of which $\wt{G}$ is a dense subgroup.}

It follows that the irreducible representation of $\wt{G}^*$
restricts irreducibly to $\wt{G}$.  Moreover, $\wh{A}$ is
evidently maximal isotropic for $\wt{G}^*$ and so this
representation can be realised on $L^2 ((\wh{A}\times
A)/(\wh{A} \times 1)) =L^2 (1 \times A)$: 

\noindent {\bf 2.5.}
{\it Every almost Heisenberg extension $\wt{G}$ of $G
= A^2$ has a distinguished irreducible representation with
natural central character, obtained by restriction from the
unique irreducible representation (having natural central
character) of the Heisenberg group $\wt{G}^*$ associated to
$\wt{G}$.  On $L^2(A)$, this representation is given by the
formulae of eqns. (3), (4), (5).}

The central extensions of $\bz^2$ relevant for the
irrational flux Peierls electron (section 4) will turn out
to be almost Heisenberg groups and the distinguished
representation described in {\bf 2.5} is the only irreducible
representation that comes into play in its quantum theory.
However, the state space of a quantum system $(\cc ,G)$
corresponding to $\alpha \in H^2 (G,\bt )$ is not an
irreducible representation of $\wt{G}_{\alpha}$, but rather
the representation of $\wt{G}_{\alpha}$ on, in general,
$L^2$ sections of a certain line bundle over $\cc$.  The
line bundle in question is that associated to $\alpha \in
H^2 (G,\bt )$ of a principal $\wh{H}^2(G,\bt)$ bundle -- by
Pontryagin duality, $\alpha$ is a character of $\wh{H}^2
(G,\bt )$.  (When 
$\cc$ is not a manifold and $G$ is not a Lie group, the
terminology is obviously meant in an algebraic sense).  It is
appropriate to name this representation as the {\it
wavefunction representation}.  Since in our applications $G$
is the translation group $\br^2$ or $\bz^2$, the state space
$\ch_{\alpha}$ is isomorphic to the space of $L^2$ functions
on $G$ itself.  Furthermore, the full symmetry groups of our
systems are the euclidean groups $E = J \vec{\times} G$ where $J$ is a
subgroup of Aut~$G$.  Before studying how the central
extensions of $E$ are related to those of $G$ and are
represented on $L^2(G) = L^2(A^2)$, we 
exhibit a decomposition of $L^2 (A^2)$ as a tensor product
of irreducible representations of (almost) Heisenberg groups
which is of great utility in all that follows.

To begin with, let $\wt{G}_{\alpha}$, $\alpha \in \ca^2
(G)$, be a central extension of any abelian group $G$, not
necessarily (almost) Heisenberg, but for which the canonical
choice of the associated 2-cocycle $\gamma_{\alpha} \in
\ca^2 (G)$ is possible.  Define an action of
$\wt{G}_{\alpha}$ on $L^2 (G) = \{ \psi, \cdots \}$ by
$$(W(g,t)_{\alpha}\psi )(h) = t\gamma_{\alpha} (g,h) \psi
(gh).\eqno (6)$$
The operators $W(g,t)_{\alpha}$ are clearly unitary on
$L^2(G)$ and, by virtue of $\gamma_{\alpha}$ being
bimultiplicative, furnish a representation of
$\wt{G}_{\alpha}$ for any $\alpha \in H^2(G,\bt )$.  If, in
addition, $\gamma_{\alpha}$ can be picked from $\ca^2 (G)$,
it is equally easy to verify that $L^2 (G)$ is a
representation of the direct product group $\wt{G}_{\alpha}
\times \wt{G}_{\alpha^{-1}}$ for the action of each factor
by eqn. (6), namely,
$$(W(g,t)_{\alpha}, (g',t')_{\alpha^{-1}})\psi )(h) =
tt'\gamma_{\alpha} (g,h)\gamma_{\alpha}(g',h')^{-1}\psi
(gg'h). \eqno (7)$$
We have used here the identity $\gamma_{\alpha^{-1}} (g,h) =
\gamma_{\alpha} (g,h)^{-1}$ and also the skewsymmetry of
$\gamma$.  Note that $\bt \times \bt \subset$ centre
($\wt{G}_{\alpha} \times \wt{G}_{\alpha^{-1}})$ operates by
$(t,t') \longmapsto tt'$ so that, as a representation of
either $\wt{G}_{\alpha}$ or $\wt{G}_{\alpha^{-1}}$, $L^2(G)$
is the lift of a  projective representation of $G$.

Noting that if $\wt{G}_{\alpha}$ is Heisenberg (almost
Heisenberg), so is $\wt{G}_{\alpha^{-1}}$, we have our key
result:

\noindent {\bf 2.6.}
{\it Let $\wt{G}_{\alpha}$ be a Heisenberg (respectively almost
Heisenberg) extension of $G$.  Then the representation of
~$\wt{G}_{\alpha} \times \wt{G}_{\alpha^{-1}}$ on $L^2(G)$
defined by eqn. (7) is irreducible.  Thus $\ch_{\alpha}
\cong L^2 (G)$ has the tensor product decomposition
$\ch_{\alpha} = V_{\alpha} \otimes V_{\alpha^{-1}}$ where
$V_{\alpha}$ is the unique (respectively distinguished) irreducible
representation of $\wt{G}_{\alpha}$ having natural central
character.} 

For the Heisenberg case, the proof is a simple extension of
the proof of the irreducibility of the representation of
$\wt{G}_{\alpha}$ on $L^2 (A)$ and will be found in [12].
For the almost Heisenberg case, we do the obvious: embed
$\wt{G}_{\alpha} \times \wt{G}_{\alpha^{-1}}$ in the
corresponding $\wt{G}_{\alpha}^* \times \wt{G}_{\alpha^{-1}}^*$
and take the irreducible representation of the latter group
on $L^2(G^*) = L^2 (A \times \wh{A})$.  This restricts to a
representation of $\wt{G}_{\alpha}\times \wt{G}_{\alpha^{-1}}$
irreducibly and, on taking Fourier transforms on $1 \times
\wh{A}$, can be written as a representation on $L^2 (A
\times A)$.

To conclude this account of the mathematical framework, we
now consider the semidirect product groups $E = J \vec{\times} G$
where $G$ as before is the (translation) group $A^2$ and $J$
is a (rotation) subgroup of Aut~$G$.  For the classification
of central extensions of $E$, we quote a general result
(for a proof, see [7]):

\noindent {\bf 2.7.}
{\it For $G$ an abelian group and $J$ a subgroup of
$Aut~G$, $H^2 (J \vec{\times} G, \bt ) =
H^2 (J,\bt ) \times H^1 (J,\wh{G}) \times H^2 (G,\bt )^J$.} 

Here $H^2 (G,\bt )^J$ is the subgroup of $H^2 (G,\bt )$
fixed pointwise by the action of $J$ and $H^1$ is the 1st
cohomology with coefficients in $\wh{G}$ considered as a
$J$-module; thus a $\wh{G}$-valued 1-cocycle on $J$ is a map
$\varphi : J \lr \wh{G}$ satisfying $\varphi (\rho \sigma ) =
\varphi (\rho ) (\rho \cdot \vp (\sigma ))$ and it is a
coboundary if there is a $\chi \in \wh{G}$ such that $\vp
(\rho ) = (\rho \cdot \chi ) \chi^{-1}$ for all $\rho,\sigma
\in J$.  The following criterion for the vanishing of $H^1$
is useful.

\noindent {\bf 2.8.}
{\it If $B$ is an abelian group divisible by 2 and
$J$ is an abelian subgroup of $Aut~B$, $H^1 (J,B)$ vanishes
whenever there exists $\rho_0 \in J$ such that $\rho_0 \cdot
\vp (\rho ) = \vp (\rho )^{-1}$ for all 1-cocycles $\vp : J
\lr B$ and all $\rho \in J$.}

For proof, we have $\vp (\rho \sigma ) = \vp (\sigma \rho)$
implying the identity $(\sigma \cdot \vp (\rho )) \vp (\rho )^{-1}
= (\rho \cdot \vp (\sigma )) \vp (\sigma )^{-1}$ from the
definition of a 1-cocycle.  Choosing $\sigma = \rho_0$ and
writing $\vp (\rho_0) = b_0 \in B$, this becomes $\vp (\rho
)^{-2} = (\rho \cdot b_0) b_0^{-1}$.  Taking square roots,
we see that $\vp$ is a coboundary.

In our applications, the conditions required for the
vanishing of $H^1 (J,\wh{G})$ will be seen to be met.  It
will also turn out that $H^2 (J,\bt ) =0$ and $H^2(G,\bt )^J
= H^2 (G,\bt )$.  Hence, in the rest of this section, we
confine attention to central extensions of $E$ of the form
$\wt{E} = J \vec{\times} \wt{G}$, with $J$ acting trivially on
$\bt \subset$ centre $\wt{G}$.  Denoting by ${\rm Aut}_0 \wt{G}$
the subgroup of ${\rm Aut}~\wt{G}$ fixing $\bt$ pointwise, $J$ is
thus a subgroup of ${\rm Aut}_0\wt{G}$.

When $\wt{G}$ is a Heisenberg group, it is a well-known fact
that every Hilbert space $V$ on which $\wt{G}$ has an
irreducible representation $V$, unique upto equivalence,
also carries a projective representation of ${\rm Aut}_0 \wt{G}$,
the metaplectic representation: If $\rho \in {\rm Aut}_0 \wt{G}$,
i.e. $\rho (g,t) = (\rho (g),t)$, and $U \mid_{\bt}$ is the
natural character, then $(g,t) \longmapsto U_{\rho} (g,t) =
U(\rho (g),t)$ is also an irreducible representation with
$U_{\rho} \mid_{\bt}$ also natural.  By the unique
representation theorem, there exist unitary operators
$O (\rho )$ on $V$ such that $U_{\rho} (g,t) = O(\rho )
U(g,t) O(\rho )^{-1}$ and $\rho \longmapsto O(\rho )$ is
clearly a representation, in general projective, of ${\rm Aut}_0
\wt{G}$ on $V$.

If $\wt{G}$ is almost Heisenberg, then ${\rm Aut}_0 \wt{G}$ is a
subgroup of ${\rm Aut}_0$ of the Heisenberg group $\wt{G}^*$.
Hence, if $V$ is a Hilbert space on which the distinguished
irreducible representation is realised, then, from the
statement {\bf 2.5}, there is a projective representation of
${\rm Aut}_0\wt{G}$ on $V$.  From this we draw the following
conclusion relevant for our purpose.  

\noindent {\bf 2.9.}
{\it Let $\wt{G}$ be a Heisenberg (almost Heisenberg)
group, $J$ a subgroup of $Aut_0\wt{G}$ such that $H^2 (J,\bt
) = 0$ and $V$ a Hilbert space on which the unique
(distinguished) irreducible representation of $\wt{G}$ is
realised.  Then there is an irreducible linear
representation of $J \vec{\times} \wt{G}$ on $V$.}

The above property carries over naturally to the wave
function representation.  When $\wt{G}_{\alpha}$ and
$\wt{G}_{\alpha^{-1}}$ are Heisenberg, all representations
of $\wt{G}_{\alpha} \times \wt{G}_{\alpha^{-1}}$ which
restrict to either factor irreducibly and nontrivially and
to the central subgroup $\bt \times \bt$ naturally are
equivalent.  And since $J$ fixes $\bt \times \bt$, it is a
subgroup of ${\rm Aut}_0 (\wt{G}_{\alpha} \times
\wt{G}_{\alpha^{-1}})$.  Similar observations apply in the
almost Heisenberg case and we have

\noindent {\bf 2.10.}
{\it $L^2 (G) = V_{\alpha} \otimes V_{\alpha^{-1}}$
is an irreducible representation of $J \vec{\times}
(\wt{G}_{\alpha} \times \wt{G}_{\alpha^{-1}})$ 
whenever $\wt{G}_{\alpha}$ is
Heisenberg or almost Heisenberg and $J \subset Aut_0
\wt{G}_{\alpha}$ has $H^2 (J,\bt ) = 0$.}

\vspace{2mm}
\noindent
{\sf 3. The Heisenberg Group of $\br^2$ and the Landau
Electron}

\vspace{2mm}
This section begins by studying the projective
representations of the real euclidean group $E (2,\br ) = SO
(2,\br )\vec{\times} \br^2$ on $L^2 (\br^2 )$ with a view to
arrive at a description of the most general quantum system
with configuration space $\br^2$ and symmetry group $E
(2,\br )$.  The central extensions of the {\it Lie algebra}
of $E (2,\br )$ were first investigated by Bargmann [6].
The general theory needed to deal with the group is given in
section 2 and is easy to apply.

First, we have $H^2 (\br^2,\bt ) = \ca^2(\br ) \cong \br$,
consisting of functions $c_r(x,y) = \exp (irx\wedge y)$ for
$x,y \in \br^2$ and $r \in \br$, all written additively.
$SO(2,\br ) = \{ \rho_{\theta} \mid 0 \leq \theta < 2\pi\}$
acts on these functions by $(\rho_{\theta}c_r)(x,y) = c_r
(\rho_{\theta} x, \rho_{\theta}y)$, $\rho_{\theta}x =
(x_1\cos \theta +x_2\sin\theta , -x_1 \sin\theta +x_2 \cos
\theta )$.  It is evident that $c_r (\rho_{\theta}
x,\rho_{\theta}y) = c_r(x,y)$; so $H^2 (\br^2,\bt
)^{SO(2,\br)} = H^2 (\br^2,\bt )$.  Also, $H^2 (SO(2,\br
),\bt ) = 0$ since $SO(2,\br )$ is 1-dimensional.  As for
the $H^1$ contribution in the statement {\bf 2.7}, the
$SO(2,\br )$ action on $\wh{\br}^2 \cong\br^2$ is $x
\longmapsto \rho_{\theta}^T x = \rho_{\theta}^{-1}x$ and it
is immediately verified that $\rho_{\theta = \pi} x = -x$.
Hence, by {\bf 2.8}, $H^1 (SO(2,\br ),\br^2)$ also vanishes
and we have

\noindent {\bf 3.1.} {\it Inequivalent
central extensions of $E(2,\br )$ by $\bt$ form
a one real parameter family of groups $\wt{E}_r(2,\br ) = SO
(2,\br ) \vec{\times} \wt{\br}_r^2$.}

The $SO(2,\br )$ action on $\wt{\br}_r^2$ is the one on
$\br^2$ extended trivially to its centre $\bt$.  Therefore, by
{\bf 2.9}, an irreducible representation of the real
Heisenberg group $\wt{\br}_r^2$, $r \neq 0$, on $V_r$ say,
is also an irreducible representation of $\wt{E}_r (2,\br
)$. For classifying all actions of $\wt{E}_r(2,\br )$ on
$V_r$, it is convenient to look at the corresponding Lie
algebra actions.  Choosing a basis $\{ L,P_1,P_2,1\}$ for
Lie $\wt{E}_{\alpha} (2,\br )$ where $L$ is the angular
momentum generating rotations, $P_1$ and $P_2$ are mutually
perpendicular momenta generating translations and 1
generates the centre, we have the Lie brackets
$$[P_1,P_2] = ir \eqno (8)$$
$$[L,P_1] = i P_2, ~~ [L,P_2] = -iP_1. \eqno (9)$$
One checks that $L+(P_1^2+P_2^2)/2r$, $r\neq 0$, has
vanishing brackets with $L,P_1$ and $P_2$ and, since $V_r$
is irreducible, is represented by a scalar $s$:
$$L = s-\frac{1}{2r} (P_1^2+P_2^2) \eqno (10)$$
for some $s \in \br$. But we know the spectrum of
$(P_1^2+P_2^2)/2r$ in $V_r$ to be $\pm \bn \mp \frac{1}{2}$
depending on the sign of $r$ (the energy spectrum of the
harmonic oscillator) and the spectrum of $L$ to be contained
in $\bz$, the characters of $SO(2,\br )$.  Hence $s \in \bz
+ \frac{1}{2}$. (Note that working with the Lie algebra of $
\wt{\br}_r^2$ is legitimate on account of $
\br^2$ being simply connected [11,17]).  Every $s$ in this
set defines a distinct set of characters of $SO(2,\br)$ and
a distinct irreducible representation of $\wt{E}_r (2,\br )$
on $V_r$.  More precisely, we have

\noindent {\bf 3.2.}
{\it Let $V_r$ be an irreducible representation space of
the real Heisenberg group $\wt{\br}^2_r$.  Then, given any
$l_0 \in \bz$, there is an irreducible representation 
of $\wt{E}_r(2,\br )$ on $V_r$ having the angular momentum
decomposition}
$$V_r = \bigoplus_{l\leq l_0} V_{r,l} ~~\mbox{\it or}~~ V_r =
\bigoplus_{l \geq l_0} V_r,l$$
{\it for $r > 0$ or $r<0$ respectively, each (one
dimensional) $V_{r,l}$, with $LV_{r,l} = lV_{r,l}$, occurring once
in the sum}.

In accordance with the general theory of section 2, a choice
for $V_r$ is $V_r = L^2 (\br )$, the space of functions of
the momentum along a fixed direction. 
Turning to the wave function representation, statements {\bf
2.6} and {\bf 2.10} have the corollary

\noindent {\bf 3.3.}
{\it For $r \neq 0$, the state space $\ch_r \cong L^2
(\br ^2)$ is the unique irreducible representation $V_r
\otimes V_{-r}$ of $\wt{\br}_r^2 \times \wt{\br}_{-r}^2$.
It is also an irreducible representation of $SO(2,\br )
\vec{\times} (\wt{\br}^2_r \times \wt{\br}_{-r}^2)$.}

The actions of $\wt{\br}_r^2$, $\wt{\br}_{-r}^2$ and
$SO(2,\br )$ on $L^2 (\br^2)$ are completely specified by the
action of the corresponding Lie algebras.  The interested
reader will find them written down and their physical
meaning discussed in [10].  We note that if $Q_1$ and $Q_2$
are noncentral basis vectors of Lie $\wt{\br}_{-r}^2$
satisfying the Lie brackets
$$[Q_1,Q_2] = -ir, \eqno (11)$$
$$[L,Q_1] = iQ_2, ~~~ [L,Q_2] = -iQ_1, \eqno (12)$$
with $[P,Q] =0$ then {\bf 3.3} has the consequence that $L$
is a polynomial in $\{ P,Q\}$.  Its form is obtained by
interpreting $s$ in eqn. (10), which is a scalar in $V_r$, as
a polynomial in $Q$.  The brackets (11) and (12) then fix
$s$ and yield 
$$L = \frac{1}{2r} (Q_1^2+Q_2^2) - \frac{1}{2r}
(P_1^2+P_2^2), \eqno (13)$$
upto an additive integer scalar.

This completes the description of the kinematics, namely the
structure of the state space, of the $E(2,\br )$-symmetric
quantum mechanics of a particle in $\br^2$.  One aspect of
the general theory not invoked so far (because it has no
significant physical role in magnetic field problems) is worthy of
passing mention: for a system $(\cc ,E)$, every
$\ch_{\alpha}$, $\alpha \in H^2 (E,\bt )$, is a
superselection sector [7,8].  In the following, we shall
refer to $\ch_{\alpha}$ for each $\alpha$ as a sector and to
$\ch_{\alpha=0}$ as the trivial sector.

In approaching the question of dynamics, i.e., in looking
for Hamiltonians to generate time evolution respecting the
symmetries of the system, the central point to keep in mind
is that there is no sense in which there is a unique
Hamiltonian valid for all sectors [7,8].  To illustrate, let
us assume that $\cc$ is a $d$-dimensional manifold which is
a homogeneous space for the (connected) Lie group $E$, $\cc
= E/R$, and let $\{ X_i\}$ be a basis for Lie $E$ adapted to
$R$, i.e., $\{ X_i \mid \dim E - \dim R < i \leq \dim E\}$
is a basis for Lie $R$, so that $\{ X_1,\ldots ,X_{d}
\}$ is a vector space basis for Lie $E/$ Lie $R$.  $\{
X_1,\ldots ,X_{d}\}$ is thus a set of mutually
perpendicular velocity vectors of the particle whose
configuration space is $\cc$.  Let $H_0$ be a nondegenerate
symmetric quadratic polynomial in the velocity vectors,
invariant under the adjoint action of $E$.  In a
representation of $E$, $H_0$ is represented by a selfadjoint
operator and is a satisfactory free Hamiltonian in the
trivial sector $\ch_0$.  To the extent that $E$-invariance
fixes the symmetric coefficients occurring in $H_0$ upto an
overall scale, 
$H_0$ is unique modulo an additive and a multiplicative
scalar [7].

In a nontrivial sector $\ch_{\alpha}$, $H_0$ is not the
correct Hamiltonian because it cannot be invariant under
$\wt{E}_{\alpha}$ (as it should be) though it is still
defined as an element of the symmetric algebra of Lie
$\wt{E}_{\alpha}$.  (These aspects are examined in detail in
[7]).  The correct $\wt{E}_{\alpha}$-invariant Hamiltonian
$H_{\alpha}$ is found as follows [7]:

\noindent {\bf 3.4.}
{\it Suppose $E$ is a connected Lie group such
that $H^2 (E,\bt )$ is in bijective correspondence with $H^2
({\rm Lie}~E, \br )$ and $R$ a subgroup of $E$ with $H^2
(R,\bt ) = 0$.  If $H_0$ is the ($E$-invariant) free
Hamiltonian of the system $(\cc = E/R, E)$, then there is
$X^{(\alpha )} \in {\rm Lie}~\wt{E}_{\alpha}$ such that
$H_{\alpha} = H_0 + X^{(\alpha )}$ is
$\wt{E}_{\alpha}$-invariant.} 

This $H_{\alpha}$ is a suitable kinetic energy in the sector
$\ch_{\alpha}$.  It does not, indeed cannot, describe free
(plane wave) motion -- there is no free motion in a
nontrivial sector.

The application of {\bf 3.4} to our system $E = E(2,\br )$,
$\cc = \br^2 = E(2,\br )/SO(2,\br )$ is immediate.  The free
Hamiltonian is of course $H_0 = (2m)^{-1}(P_1^2+P_2^2)$.  For $r \neq
0$, $\wt{E}_r (2,\br )$ does not leave $H_0$ invariant.  But there is a
unique $X^{(r)} = rm^{-1} L \in {\rm Lie}~\wt{E}_r (2,\br )$
such that
$$H_r = H_0 + X^{(r)} = \frac{1}{2m}(P_1^2+P_2^2) +
\frac{r}{m} L \eqno (14)$$
is fixed by $\wt{E}_r (2,\br )$.  The expression (13) for
$L$ simplifies this to 
$$H_r = \frac{1}{2m} (Q_1^2+Q_2^2), \eqno (15)$$
i.e., on $\ch_r = V_r \otimes V_{-r}$, $H_r$ is the operator
$1 \otimes (Q_1^2+Q_2^2)/2m$.  The spectrum of $H_r$ is
$(rm^{-1}) (\bn -\frac{1}{2})$ with multiplicity one in
$V_{-r}$; on the whole of $\ch_r$, the eigenspaces of $H_r$
are $V_r$ for every eigenvalue and for all $r \neq 0$.  The
spectrum thus matches the energy eigenvalues and degeneracies 
of the Landau electron moving in a magnetic field $B$ on
identifying the nonzero real number $r$ with $eB$.

We may now use $H_r$ to write down the Heisenberg equation
of motion for any operator on $\ch_r$.  For $Q$ ($P$ and $J$
are automatically conserved) we find 
$$\frac{dQ_1}{dt} = i [H_r,Q_1] = - \frac{r}{m} Q_2, ~~ 
\frac{dQ_2}{dt} = \frac{r}{m} Q_1. \eqno (16)$$
These constitute the Lorentz force equation if $Q$ is taken
to be proportional to the velocity $v$ (with $r=eB$); the
velocity dependence of energy then fixes $Q = mv$.

The results of this section have established our claim that
the projective representation theory of $E(2,\br )$ on $L^2
(\br^2)$ {\it is} the quantum theory of the Landau
electron.  The treatment may appear somewhat abstract, but
has the great advantage of dispensing with all but the one
essential physical parameter, namely the magnetic field.
Many conceptual issues are thereby clarified especially the
origin and (lack of) significance of gauges and gauge
transformations, the origin of degeneracies, the fact that
velocity is not proportional to momentum, ambiguities in the
classical mechanics of (electro) magnetic problems, etc.  A
fuller account of these aspects will be found in [10].

\vspace{2mm}
\noindent
{\sf 4. The $\bz^2$ Peierls Electron for Irrational Fluxes}

\vspace{2mm}
In this section we take up a particle moving on the
infinite planar square lattice, $\cc = \bz^2 \subset
\br^2$, and having the discrete euclidean group $E (2,\bz )
= \bz/4 \vec{\times} \bz^2$ as its group of symmetries.  Denoting
by $\zeta$ the generator of $\bz /4$ corresponding to an
anticlockwise rotation by the angle $\pi/2$, the
action of $\bz /4$ on $\bz^2$ is $\zeta \cdot (m_1,m_2) =
(-m_2,m_1)$, $m = (m_1,m_2) \in \bz^2$ (written
additively).

A general bicharacter $b$ on $\bz^2$ is a function $b(m,n) =
\exp (i\theta_1m_1n_2 + i\theta_2 m_2n_1 + i\theta_3 m_1n_1
+ i \theta_4m_2 n_2)$ with $\theta_i \in [0,2\pi )$.  If $b$
is in addition alternating, then $\theta_1 + \theta_2 =
\theta_3 = \theta_4 = 0$ (mod $2\pi$).  Hence
$$H^2 (\bz^2,\bt ) = \ca^2 (\bz^2) = \{ c_{\theta} : \bz^2
\lr \bt \mid c_{\theta} (m,n) = e^{i\theta m \wedge n} \}
\cong \bt .$$ 
The action of $\bz /4$ on $\ca^2(\bz^2)$: $\zeta \cdot c_{\theta}
(m,n) = c_{\theta} ~(\zeta \cdot m,\zeta \cdot n)$ leaves every
$c_{\theta}$ fixed.  Also, $H^2 (\bz /4,\bt )$ and $H^1
(\bz /4,\wh{\bz}^2) = H^1(\bz /4,\bt^2)$ both vanish, the
latter because $\rho_0 = \zeta^2$ meets the requirements of
{\bf 2.8}.  Hence, as in the real case, central extensions
of $E(2,\bz )$ and $\bz^2$ are in 1-1 correspondence:

\noindent {\bf 4.1.}
{\it Every central extension of $E(2,\bz )$ by
$\bt$ is of the form $\wt{E} (2,\bz ) = \bz/4 \vec{\times}
\wt{\bz}^2$, where $\bz /4$ acts on $\bt \subset \wt{\bz}^2$
trivially.  Inequivalent central extensions of $\bz^2$ and
hence of $E(2,\bz )$ are parametrised by an angle $\theta
\in [0,2\pi )$.}

The relationship of the projective representations of
$E(2,\bz )$ and of $\bz^2$ with the quantum mechanics of the
$\bz^2$ Peierls electron becomes manifest already at this
point.  Reverting to the Landau electron briefly, we observe
that the commutator of translations through $x$ and $y$ in
$\wt{\br}^2_r$ has the physical meaning
$$c_r(x,y) = e^{irx\wedge y} = e^{ie\Phi (x,y)}$$
where $\Phi$ the magnetic flux through a parallelogram
having $x,y \in \br^2$ as adjacent sides.  Embedding $\bz^2$
in $\br^2$ as the lattice generated by the vectors
$(\xi_1,0)$, $(0,\xi_2) \in \br^2$ and equating the
commutator of $(1,0)$ and $(0,1)$ in $\wt{\bz}^2_{\theta}$
to that of $(\xi_1,0)$ and $(0,\xi_2) $ in $\wt{\br}^2_r$,
we get 
$$e^{i\theta} = e^{ieB\xi_1\xi_2} = e^{ie\Phi} =
e^{i2\pi\Phi/\Phi_0} \eqno (17)$$
where the constant $\Phi$ is the flux through the plaquette
bounded by the generators and $\Phi_0 = 2\pi/e$ is the flux
quantum.  Choosing units in which $\Phi_0=1$, we may
identify $\theta$ with $2\pi \Phi$.

>From the above we conclude:

\noindent {\bf 4.2.}
{\it The quantum mechanics of the $\bz^2$ Peierls
electron is fully determined by, and is periodic in, one
physical parameter, namely the flux per plaquette $\Phi$,
with period equal to one flux quantum.}

In particular, when the flux is integral, the motion of the
particle is free hopping motion.  It is also evident that
the field $B$ itself is totally irrelevant.

The numerical work of Hofstadter [18] on the spectrum of the
Harper Hamiltonian has demonstrated, very graphically, that
it depends qualitatively on whether the flux is
rational or irrational as a multiple of the flux quantum.
Such differences are reflections of the differences in the
structure and representation theory of $\wt{\bz}_{\theta}^2$
for rational and irrational values of $\theta /2\pi$.  The
commutator function $c_{\theta}$ defines the map
$\mu_{\theta}: \bz^2 \lr \bt^2$ by $\mu_{\theta}(m_1,m_2) =
(\exp (-i\theta m_2),\exp (i\theta m_1))$.  If $\theta$ is
rational, $\theta = 2\pi \nu/N$ with $\nu$ and $N$ coprime,
then $\mu_{\theta}$ has kernel $(N\bz)^2$.  This rational
flux case of the Peierls electron will be the subject of a
sequel; in the rest of this paper, we deal with the case of
irrational fluxes.

When $\theta$ is irrational, $\mu_{\theta}$ is an injective
map and its image is dense in $\bt^2$.  Hence 
$\wt{\bz}_{\theta}^2$ is an almost Heisenberg group.
Moreover, $b^2_{\theta} \neq 1$ for any irrational
bicharacter $b_{\theta}$ and hence we can choose the
2-cocycle corresponding to $\wt{\bz}_{\theta}^2$
canonically, as the skew symmetric square root of
$c_{\theta}$ (see remarks following {\bf 2.1}):
$$\gamma_{\theta} (m,n) = e^{\frac{1}{2} i\theta m\wedge n} \eqno
(18)$$

The distinguished irreducible representation and the wave
function representation of $\wt{\bz}_{\theta}^2$ can now be
characterised.  Denoting the
noncentral generators of $\wt{\bz}_{\theta}^2$ by $p_1 =
(1,0,1)$, $p_2 = (0,1,1)$  ($\bz^2$ is written additively and
$\bt$ multiplicatively) and the corresponding generators
of $\wt{\bz}^2_{-\theta}$ by $q_1,q_2$ satisfying
$$p_1p_2p_1^{-1}p_2^{-1} = e^{i\theta}, ~~~~
q_1q_2q_1^{-1}q_2^{-1} = e^{-i\theta}, \eqno (19)$$
we have

\noindent {\bf 4.3.}
{\it For any irrational $\theta/2\pi ,\wt{\bz}_{\theta}^2$
has an irreducible representation on $L^2 (\bz )$ given by
$$(U(p_1)f) (m) = e^{i\theta m} f(m), ~~~~ (U(p_2)f)(m) =
f(m+1).$$
By {\bf 2.9} this extends also to an irreducible representation of
$\wt{E}_{\theta} (2,\bz )$.}

\noindent {\bf 4.4.}
{\it There is an irreducible representation of
$\wt{\bz}_{\theta}^2 \times \wt{\bz}^2_{-\theta}$ and $\bz /4
\vec{\times} (\wt{Z}_{\theta}^2 \times \bz_{-\theta}^2)$ on $L^2
(\bz^2)$ given by}
$$\begin{array}{l}
(W(p_1)\psi ) (m_1,m_2) = e^{\frac{1}{2}im_2 \theta} \psi
(m_1+1,m_2), \\
(W(p_2)\psi ) (m_1,m_2) = e^{-\frac{1}{2}im_1 \theta} \psi
(m_1,m_2+1), \\
(W(q_1)\psi ) (m_1,m_2) = e^{-\frac{1}{2}im_2 \theta} \psi
(m_1+1,m_2), \\
(W(q_2)\psi ) (m_1,m_2) = e^{\frac{1}{2}im_1 \theta} \psi
(m_1,m_2+1), \\
\end{array}$$
$$(W(\zeta )\psi ) (m_1,m_2) = \psi (-m_2,m_1).$$
{\it If $V_{\theta}$  is the Hilbert space of the
distinguished irreducible representation of
$\wt{\bz}_{\theta}^2$, $L^2 (\bz^2)$ is isomorphic to
$V_{\theta} \otimes V_{-\theta}.$}

The second part of {\bf 4.4} is equivalent to the statement
that $\wt{\bz}_{\theta}^2 \times \wt{\bz}_{-\theta}^2$ has a
unique irreducible representation upto equivalence with the
property that its restriction to either factor is the
distinguished irreducible representation.

This is the key representation-theoretic foundation of the
quantum mechanics of the $\bz^2$ Peierls electron for
irrational fluxes.  The space $L^2 (\bz^2)$ is the state
space $\ch_{\theta}$ for the action given above of
$\wt{E}_{\theta} (2,\bz )$.  Similar to the Landau case,
only the subgroup $\wt{E}_{\theta} (2,\bz )$ of $\bz /4
\vec{\times} (\wt{\bz}_{\theta}^2 \times \wt{\bz}^2_{-\theta})$ is
directly related to the physical symmetries; the group
$\wt{\bz}_{-\theta}^2$ just parametrises the possible
inequivalent actions of $\wt{E}_{\theta} (2,\bz )$ (or the
multiplicity of the distinguished representation of
$\wt{\bz}_{\theta}^2 )$ in $L^2 (\bz^2)$.

The functions $\psi \in L^2 (\bz^2)$ are the wave
functions.  The action of $p_1,p_2,q_1$ and $q_2$ given in
{\bf 4.4} results from the canonical choice of the 2-cocycle
$\gamma_{\theta}$ as an element of $\ca^2 (\bz^2)$, eqn. (18).
We have the freedom to modify $\gamma_{\theta}$ by a coboundary
-- the symmetric bimultiplicative function $\exp (i\phi
(m_1n_2 + m_2n_1))$ -- without changing the structure of the
groups $\wt{\bz}_{\theta}^2$ and $\wt{E}_{\theta} (2,\bz
)$.  The representation $W$ gets modified thereby to 
$$W' (p_1 ) = e^{i\phi m_2} W(p_1), ~~ W(p_2 ) = e^{i\phi
m_1} W(p_2),$$
etc.  But $W$ and $W'$ are both restrictions to a dense
subgroup of irreducible representations $W^*$ and $W^{*'}$
of the group $\wt{\bz}_{\theta}^{2*} \times
\wt{\bz}_{-\theta}^{2*}$ in which each factor is a
Heisenberg extension of $\bz \times \bt$.  Consequently
$W^*$ and $W^{*'}$ are equivalent and hence so are $W$ and
$W'$: $W' (m,n) =SW (m,n)S^{-1}$ for some unitary operator
$S$ on $\ch_{\theta}$.  The operators representing
translations in $\ch_{\theta}$ for a given irrational flux
form a one (angle) parameter family and are related among
themselves by unitary operators $S(\phi )$.  Physically, $S$
is a gauge transformation and the (unitary) equivalent
representations of $\bz /4\vec{\times} (\wt{\bz}_{\theta}^2
\times \wt{\bz}_{-\theta}^2)$, parametrised by $\phi \in
[0,2\pi )$, are gauge-equivalent.  The justification for
this assertion is the corresponding phenomenon in the Landau
case where, infinitesimally, $S$ arises from an
$\br^2$-valued 2-coboundary added to a connection on a
$U(1)$ bundle, the vector potential, without changing the
curvature [10].  In any case, our treatment bypasses all
questions related to gauges, except in so far as the action
of the symmetry group on wave functions is desired in an
explicit form.  

Our final task is the determination of the Hamiltonian(s)
$H_{\theta}$ governing time evolution in the sector
$\ch_{\theta}$.  As $\ch_{\theta}$ is irreducible under the
action of $\wt{\bz}_{\theta}^2 \times \wt{\bz}_{-\theta}^2$,
$H_{\theta}$ as an operator on $\ch_{\theta}$ is a
selfadjoint element of the algebra of operators representing
this group.  $\wt{E}_{\theta} (2,\bz )$ acts on the operator
algebra by conjugation by unitary operators; $H_{\theta}$
must be invariant under this action in order to preserve the
symmetries under time evolution.

Since we no longer have at our disposal infinitesimal
operators representing momenta and velocities, the
counterpart to the procedure followed
for the Landau electron is to find the subalgebra of the
group algebra $\bc [\wt{\bz}^{2}_{\theta}\times\wt{\bz}_{-\theta}^{2}]$
satisfying the two conditions of selfadjointness and
pointwise invariance under $\wt{E}_{\theta}(2, \bz)$. Given the
commutators of eqns. (19) and
$$\zeta p_{1}\zeta^{-1}=p_{2},~~~~~\zeta
p_{2}\zeta^{-1}=p_{1}^{-1}, \eqno (20)$$
$$\zeta q_{1}\zeta^{-1}=q_{2},~~~~~\zeta
q_{2}\zeta^{-1}=q_{1}^{-1}, \eqno(21)$$
this is a straightforward computation.\\

Confine attention first to $\bc [\wt{\bz}^{2}_{\theta}].$ An
element of this algebra has the general form
$$\Omega_{\theta}=\sum w_{j_1j_2}p_{1}^{j_1}p_{2}^{j_2} \eqno
(22)$$
where the sum is over $j_{1},j_{2}\in\bz$, using the
fact that $p_{1}p_{2}p_{1}^{-1}p_{2}^{-1}$ is in the centre of 
 $\wt{\bz}_{\theta}^{2}$, and hence is a scalar. Then
$$p_{1}\Omega_{\theta}p_{1}^{-1}=\sum
w_{j_{1}j_{2}}e^{ij_{2}\theta}p_{1}^{j_1}p_{2}^{j_2},$$
$$p_{2}\Omega_{\theta}p_{2}^{-1}=\sum
w_{j_{1}j_{2}}e^{-ij_{1}\theta}p_{1}^{j_1}p_{2}^{j_2}.$$
So for $\wt{\bz}_{\theta}^{2}$ to fix $\Omega _{\theta}$, we have the
conditions
$$\sum w_{j_{1}j_{2}}(1-e^{ij_{2}\theta})p_{1}^{j_1}p_{2}^{j_2}=0,$$
$$\sum
w_{j_{1}j_{2}}(1-e^{-ij_{1}\theta})p_{1}^{j_1}p_{2}^{j_2}=0.$$
Since the only relation among the generators of
$\wt{\bz}_{\theta}^{2}$ is eqn.~(19), $p_{1}$ and $p_{2}$ generate
$\wt{\bz}_{\theta}^{2}$ freely modulo its centre. Hence the above
equations hold only if the coefficients of
$p_{1}^{j_1}p^{j_2}$ vanish for all nonzero $j_{1},j_{2}\in
\bz$. But since $\theta$ is irrational, $\exp(ij\theta )\neq 1$ for any
$j\neq 0$, leaving $w_{00}$ as the only nonzero
coefficient: the centre of $C[\wt{\bz}_{\theta}^{2}]$ is
generated by the centre of 
$\wt{\bz}_{\theta}^{2}$ and hence consists of scalars. It follows
that the subalgebra of $\bc[\wt{\bz}_{\theta}^{2}\times
\wt{\bz}_{-\theta}^{2}]$ fixed pointwise by $\wt{\bz}_{\theta}^{2}$ is
$\bc [\wt{\bz}_{-\theta}^2]$ consisting of elements
$\Omega_{-\theta}$ obtained from eqn. (22) by replacing $p$
by $q$.  On this we have the $\bz /4$ action given by
eqn. (21) implying that elements of $\bc
[\wt{\bz}_{\theta}^2 \times \wt{\bz}_{-\theta}^2]$ invariant
under $\wt{E}_{\theta} (2,\bz )$ are of the form
$$\Omega _{-\theta}=\sum
w_{j_{1}j_{2}}(q_{1}^{j_1}q_{2}^{j_2}+q_{2}^{j_1}q_1^{-j_2}
+q_{1}^{-j_1}q_{2}^{-j_2}+q_{2}^{-j_1}q_{1}^{j_2})$$
for arbitrary complex coefficients  $w_{j_{1}j_{2}}.$

The requirement that $\Omega _{-\theta}$ be selfadjoint when
$q_1$ and $q_2$ are represented by unitary operators imposes
the final condition
$$\sum
w_{j_{1}j_{2}}(q_{1}^{j_1}q_{2}^{j_2}+e^{-ij_1,j_{2}\theta}
q_{1}^{-j_2}q_{2}^{j_1}+q_{1}^{-j_1}q_2^{-j_2}+e^{ij_1j_{2}\theta}q_{1}^{j_2}
q_{2}^{-j_1})$$
$$=\sum \overline{w}_{j_{1}j_{2}}(e^{-ij_1,j_{2}\theta}
q_{1}^{-j_2}q_{2}^{-j_{1}}
+q_{1}^{j_2}q_{2}^{-j_1}+e^{ij,j_{2}\theta}q_{1}^{j_1}q_2^{j_2}
+q_{1}^{-j_2}q_{2}^{j_1}),$$
where we have ordered factors using the commutator of eqn.
(19). But, again, since $q_1$ and $q_2$ are free generators
of $\wt{\bz}_{-\theta}^{2}$ modulo its centre, the above relation
can hold only if the coefficient of every monomial  
$q_{1}^{j_1}q_{2}^{j_2},j_{1}\neq 0,j_{2}\neq 0,$ is the
same on both sides. Hence we must have
$$w_{j_{1}j_{2}}=\overline{w}_{j_{1}j_{2}}e^{ij_{1}j_{2}\theta},~~\overline{w}_{j_1j_2}=\overline{w}_{j_{1}j_{2}}e^{-ij_1j_{2}\theta}$$
simultaneously for all $j_{1},j_{2}\neq 0$. This is possible
only if exp $(2ij_{1}j_{2}\theta)=1$ whenever
$w_{j_{1}j_{2}}\neq 0$ and, since $\theta$ is irrational, we have
$w_{j,j_2}=0$ unless $j_{1}=0$ or $j_{2}=0$. The surviving
terms in $\Omega_{-\theta}$ are therefore
$$\Omega_{-\theta}=\sum_{j\in\bz^+}(w_{j}+\overline{w}_{j})
(q_{1}^{j}+q_{1}^{-j}+q_{2}^{j}+q_{2}^{-j}). \eqno (23)$$
We have thus demonstrated 

\noindent
{\bf 4.4.} 
{\it The most general selfadjoint
$\wt{E}_{\theta}(2,\bz)$-invariant element of
$\bc[\wt{\bz}_{\theta}^{2}\times\wt{\bz}_{-\theta}^{2}]$ is given by
eqn.(22) with arbitrary real coefficients.}

According to the remarks earlier in this section, all
possible Hamiltonians are obtained by restricting the sum in
eqn.(23) to a finite number of terms. The Harper Hamiltonian
results on keeping only the nearest neighbour terms, $j=1$,
upto an additive constant $w_{0}+\overline{w}_0$ of no
significance and a multiplicative constant
$w_{1}+\overline{w}_1$ which is just a scale. We
summarise:

\noindent
{\bf 4.5.} {\it For the Peierls electron on $\bz^{2}$ in
irrational flux per plaquette $\Phi =\theta/2 \pi$},\\
i)  {\it the state space has the structure
$\ch_{\theta}=V_{\theta}\otimes V_{-\theta}$, where $V_{\theta}$ is the
distinguished irreducible representation of the almost
Heisenberg group $\wt{\bz}_{\theta}^{2}$}\\
ii)  {\it the $\wt{E}_{\theta}(2,\bz)$- invariant
Hamiltonian restricted to  the nearest neighbour term is,
upto an additive and a multiplicative constant, of the form
$1\otimes H_{\theta}$, where $H_{\theta}$ acting on $V_{-\theta}$ is
the Harper Hamiltonian;}\\
iii)  {\it consequently, every energy eigenvalue has
$V_\theta$ as degeneracy subspace.}\\
As for the Landau electron, the infinite degeneracy of
energy levels is a direct reflection of translation and
euclidean invariance.

\vspace{2mm}
\noindent
{\sf Acknowledgements}

\vspace{2mm}
Discussions with M.S. Raghunathan, S. Ramanan and R.R. Simha of the Tata 
Institute of Fundamental Research, Mumbai and V.S. Sunder of the 
Institute of Mathematical Sciences, Chennai have been invaluable in the 
course of this work.  The School of Mathematics of TIFR is thanked for 
its hospitality and the IMSc for the use of its facilities.

\vspace{2mm}
\noindent
{\sf References}
\begin{enumerate}
\item
Peierls, R.E.: Z.  Phys {\bf 80}, 763 (1933).
\item
Lieb, E. and Loss, M.:  Duke Math. J. {\bf 71}, 335 (1993).
\item
Harper, P.G.: Proc. Phys. Soc. {\bf A 68}, 874 (1995).
\item Divakaran, P.P.: Heisenberg Groups in the
Theory of the Lattice Peierls Electron: The Rational Flux
Case (Paper in preparation).
\item
Wigner, E.P.:{\it Group Theory and its Applications to
the Quantum Mechanics of Atomic Spectra}, New York: Academic
Press, 1959.
\item
Bargmann, V.: Ann Math. {\bf 59}, 1 (1954).
\item
Divakaran, P P.: Rev. Math. Phys.{\bf 6}, 167 (1994).
\item
Divakaran, P P.:Phys. Rev. Lett. {\bf 79}, 2159 (1997).
\item
Divakaran, P.P.: in preparation.
\item
Divakaran, P.P and Rajagopal, A.K.: Int. J. Mod. Phys. B{\bf 9},
261(1995).
\item
Raghunathan, M.S.: Rev. Math. Phys. {\bf 6}, 207 (1994).
\item
Mumford, D.,Nori, M and Norman, P.:{\it Tata Lectures on
Theta III}, Boston, Basel,  Berlin: Birkh\"auser, 1991.
\item
Boon, M.H.: J.Math. Phys. {\bf 13},1268 (1972).
\item
Bellissard, J.: in {\it Operator Algebras and Applications},
Vol. 2, eds. Evans, D.E. and Takesaki, M., Cambridge:
Cambridge University Press, 1988. 
\item
Davidson, K.R.: {\it $C^*$-Algebra by Example}, Delhi : Hindustan
Book Agency, 1996.
\item
Prasad, G. and Raghunathan, M.S.: Invent. Math. {\bf 92},
645(1998).
\item
Varadarajan, V.S.: {\it Geometry of Quantum Theory}, Vol II, New
York: Van Nostrand Reinhold, 1970.
\item
Hofstadter, D.R.: Phys. Rev.{\bf B 14}, 2239(1976). 
\end{enumerate}
\end{document}